# Cultural Bias in Explainable AI Research: A Systematic Analysis

[This is a penultimate draft of a paper submitted to *Journal of Artificial Intelligence Research*]

**Uwe Peters**                                                        U.PETERS@UU.NL
*Department of Philosophy, Utrecht University*
*3512 BL Utrecht, The Netherlands*

**Mary Carman**                                           MARY.CARMAN@WITS.AC.ZA
*Department of Philosophy, University of the Witwatersrand*
*2050 Johannesburg, South Africa*

## Abstract

For synergistic interactions between humans and artificial intelligence (AI) systems, AI outputs often need to be explainable to people. Explainable AI (XAI) systems are commonly tested in human user studies. However, whether XAI researchers consider potential cultural differences in human explanatory needs remains unexplored. We highlight psychological research that found significant differences in human explanations between many people from Western, commonly individualist countries and people from non-Western, often collectivist countries. We argue that XAI research currently overlooks these variations and that many popular XAI designs implicitly and problematically assume that Western explanatory needs are shared cross-culturally. Additionally, we systematically reviewed over 200 XAI user studies and found that most studies did not consider relevant cultural variations, sampled only Western populations, but drew conclusions about human-XAI interactions more generally. We also analyzed over 30 literature reviews of XAI studies. Most reviews did not mention cultural differences in explanatory needs or flag overly broad cross-cultural extrapolations of XAI user study results. Combined, our analyses provide evidence of a cultural bias toward Western populations in XAI research, highlighting an important knowledge gap regarding how culturally diverse users may respond to widely used XAI systems that future work can and should address.

## 1. Introduction

To combine the strengths and mitigate the limitations of human intelligence and AI, increasingly more hybrid human-AI (HHAI) systems (e.g., AI-assisted human experts) are being developed and used (e.g., for clinical decision-making) (Chen et al., 2020). Successful HHAI systems involve and depend on close proactive collaborations, trust, and mutual understandability between humans and AI systems (Bansal et al., 2019). Yet, many of the AI systems that are now frequently used in high-stakes decision-making domains are opaque, i.e., they operate in ways too computationally complex even for AI developers to fully understand (Burrell, 2016). This opacity raises questions about these systems' trustworthiness and can undermine successful HHAI collaborations: If the humans that are part of human-AI hybrid systems cannot understand why an AI produces the output it does, they may lack meaningful control over it in their collaboration with the model (Akata et al., 2020). AI explainability is thus vital for human-AI interactions.

One main approach to dealing with this challenge is to equip AI systems with XAI models developed to make opaque systems' outputs understandable to humans (Arrieta et al., 2020). However, the ability to understand the explanations that XAI systems produce may differ between individuals (Wang & Yin, 2021), and it has been noted that XAI designers frequently adopt a one-size-fits-all approach, suiting primarily only AI experts (Ehsan et al., 2021). This may result in XAI systems that leave many AI users' explanatory needs in human-AI interactions unaddressed and that potentially operate in ways at odds with, for instance, the EU's General Data Protection Regulation (Casey et al., 2019).

Apart from interpersonal differences in expertise, *culture*, i.e., the set of attitudes, values, beliefs, and behaviors shared by a group of people and communicated from one generation to the next (Matsumoto, 1996), may also significantly influence what explanations people expect or prefer from AI systems thus affecting human-AI collaborations. The importance of cultural differences has been noted in several areas of AI research including AI ethics (e.g., people's responding to moral dilemmas faced by autonomous vehicles; Awad et al., 2018) and calls for greater cultural inclusivity in AI developments and applications are increasing (Carman & Rosman, 2021; Linxen et al. 2021; Okolo et al., 2022). However, it remains unclear to what extent there are XAI-relevant cultural variations in explanatory needs. Several AI review papers have drawn XAI researchers' attention to psychological findings on human explanations (Abdul et al., 2018; Miller, 2019). But they have not considered empirical work on cultural differences in human explanatory needs, leaving it unclear whether there are XAI-relevant cultural variations and what they would be.

A related concern is that studies in the behavioral sciences, including the fields of human-computer interaction (HCI) and human-robot interaction (HRI), found that many researchers predominantly tested only individuals from Western, educated, industrialized, rich, and democratic (WEIRD) countries, even though WEIRD people comprise only 12% of the world population (Rad et al., 2018; Linxen et al., 2021; Seaborn et al., 2023). The field of XAI might have taken countermeasures and be less affected by WEIRD sampling. However, it has not been investigated whether that is so. In a recent systematic review focusing on XAI research in the Global South, Okolo et al. (2022) found only three XAI papers that engaged with or involved people from communities in the Global South. But the authors did not examine to what extent XAI studies outside the Global South may nevertheless be culturally diverse.

More importantly, while several recent studies report that WEIRD sampling may severely limit the generalizability of HCI or HRI studies (Linxen et al., 2021; Seaborn et al., 2023), these studies do not yet control for the point that studies sampling only individuals from one kind of population can be unproblematic even if there are relevant cultural differences. After all, researchers may tailor their conclusions to their specific sample or study population, making clear that other populations remain to be explored. WEIRD sampling may only become questionable when findings are presented as if they apply beyond these populations and researchers produce 'hasty generalizations', i.e., conclusions whose scope is broader than warranted by the evidence and justification provided by the researchers (Peters & Lemeire, 2023). Relatedly, we recently found that the scope of generalizations in many XAI user studies was only poorly correlated with the size of the studies' samples suggesting that hasty, overly broad extrapolations may have been common (Peters & Carman, 2023).

However, no prior corpus analysis has investigated how broadly results are generalized across cultures in XAI user studies, leaving a significant gap in the previous work that highlights problems related to WEIRD sampling. Hasty generalizations of study results may obscure cultural variations in people's XAI needs and increase the risk that large parts of the world population are overlooked in the development of XAI and HHAI systems. Analyzing XAI user studies for hasty generalizations is therefore vital.

Here, we aim to fill the research lacunas just outlined. We offer three main contributions. First, by drawing on existing psychological studies, we argue that many popular XAI models are likely better aligned with the explanatory needs and preferences that were found in people from typically individualist, commonly WEIRD cultures than with those that were found in people from typically collectivist, commonly non-WEIRD cultures. We outline a range of cultural differences that may affect many people's perception of XAI outputs, making them relevant for research on human-AI collaborations. Second, we analyzed an extensive corpus of over 200 XAI user studies to examine whether they indicate awareness of cultural variations in explanations, have diverse samples, or avoid overgeneralizing their results (e.g., to non-WEIRD populations that were not tested). We found that most of these studies failed on all three counts. Finally, to see whether these problems have been noticed in XAI user research, we also systematically analyzed more than 30 literature reviews of XAI user studies. Most reviews, too, did not indicate any awareness of relevant cultural variations in people's explanatory needs. Nor did they mention the problems of WEIRD sampling and hasty, overly broad generalizations of results to non-WEIRD individuals in XAI user studies. Our analyses therefore provide evidence of both a significant cultural bias toward WEIRD populations and a knowledge gap on whether popular XAI models' outputs are satisfactory across cultures. We end with a set of recommendations to culturally diversify XAI user studies.

## 2. Explainable AI Focusing on Internal Factors Risks Overlooking Collectivist Cultures

Two broad categories of XAI techniques are often distinguished: transparent models, which are strictly interpretable because of their relatively simple structure (e.g., linear and logistic regression models, short decision trees), and post-hoc systems, which may either directly access or infer factors causally contributing to an opaque model's decisions after its training (Arrieta et al., 2020). Post-hoc models currently dominate XAI designs for lay-users (Taylor & Taylor, 2021). Their outputs may be visual (e.g., saliency maps), numerical (e.g., importance scores), or textual (e.g., feature reports), and generally cite factors that are internal to an opaque model and determinative of its decision (Arrieta et al., 2020). In that sense, post-hoc XAI outputs are often thought to be analogous to the human way of explaining decisions in terms of internal mental states (belief, desires, etc.) (Adadi & Berrada, 2018; Zerilli, 2022) and frequently contain mentalistic notions ('being confident', 'think', 'know'). Table 1 presents examples.

| |
|---|
| (1) XAI: "I am $C x($ $)$ confident that $y$ will be correct based on $|S|$ past cases deemed similar to $x$." (Waa et al., 2020, p. 4) |
| (2) XAI: "Here is why the classifier thinks so [presentation of (e.g.) a decision tree]." (Yang et al., 2020, p. 1) |
| (3) XAI: "Why this exercise? Wiski thinks your current level matches that of this exercise!" (Ooge et al., 2022, p. 3) |
| (4) XAI: "ShapeBot knows this is a [AI output] because ShapeBot realizes [decision factors]." (Zhang et al. 2022, p. 10) |

Table 1: Four examples of internalist XAI outputs from XAI user studies

Post-hoc XAI systems producing such internalist explanations have been criticized for being "algorithm-centered", as they tend to ignore the social context in which AI systems operate (Ehsan et al., 2021). Yet, many AI researchers now hold that for lay-users, XAI systems *should* provide explanations that cite internal states that are viewed as analogues to human beliefs or desires because they are shorter, easier to understand, and people expect such explanations (De Graaf & Malle, 2017; Zerilli et al., 2019). Correspondingly, a "significant body of work in XAI aims to explain ML [machine learning] systems by reducing their operations to a form that is amenable to belief-desire representation" and so "intentional stance" interpretations (Zerilli, 2022, p. 2). Hence, many currently popular XAI designs for lay-users rest on the assumption that people in general prefer internalist explanations of behavior, i.e., explanations invoking an agent's intentional, inner states.

However, none of the just cited papers that argue that XAI systems should provide such (intentional stance) explanations have so far reflected on whether this kind of explanation is equally used and accepted across all cultures. This is problematic, as the explanations that people prefer for a given decision or action are unlikely to be uniform cross-culturally. To illustrate this point, we will focus on internalist explanations and variations between *individualist* cultures, where a person's self is often viewed as a discrete entity independent of others, and *collectivist* cultures, where a person's self is often viewed as interdependent with others (Hampton & Varnum, 2020). While differences between these two cultures are not limited to particular regions (Fatehi et al., 2020), and people within a country are usually highly heterogeneous preventing a clear demarcation of cultures by countries (Oyserman et al., 2002), several recent studies found that WEIRD countries (e.g., the USA) were predominantly individualist whereas non-WEIRD countries (e.g., China) were predominantly collectivist cultures (Klein et al., 2018; Pelham et al., 2022). Figure 1 visualizes this evidence on the link between cultures and countries.

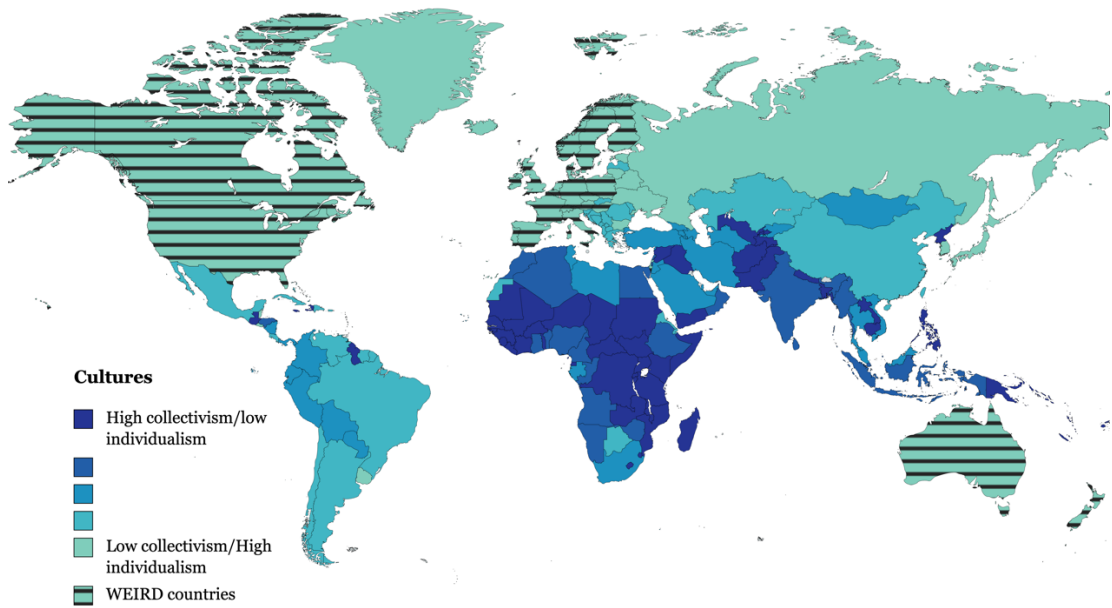

Figure 1: World map displaying the geographical distribution of collectivist and individualist cultures; horizontal stripes indicate WEIRD countries (map was self-created using MapChart).

The overlap between individualist cultures and WEIRD countries and collectivist cultures and non-WEIRD countries, respectively, is important here because psychological studies on human explanations consistently found that while participants from individualist, typically WEIRD cultures such as the USA, did tend to explain behavior primarily in terms of an agent's internal mental features (e.g., attitudes, character, or beliefs), participants from collectivist, commonly non-WEIRD cultures such as India, Korea, Saudi Arabia, and China, instead preferentially explained behavior in terms of *external* factors including social norms, task difficulty, or economic circumstances (Miller, 1984; Cha & Nam, 1985; Al-Zahrani & Kaplowitz, 1993; Lillard, 1998). To illustrate the difference, suppose an observer sees, for instance, a nurse assisting a patient in trouble, or a man robbing a bank. If the observer has an externalist focus in their behavior explanation, they may hold that the nurse acts that way because she has the social role to look after patients, and the man committed the crime because of economic hardship, respectively, rather than internal factors such as desires or beliefs.

Studies exploring such differences in externalist vs. internalist explanations found that many people from non-Western populations (i.e., Asian-Australian, Chinese-Malaysian, Filipino, Japanese, Mexican) more strongly endorsed ideas that suggested that internal traits did "not describe a person as well as roles or duties do, and that trait-related behavior changes from situation to situation" (Henrich et al., 2010, p. 12). Correspondingly, in Pacific societies, many people were found to be under the expectation to "refrain from speculating (at least publicly) about what others may be thinking" (Robbins & Rumsey, 2008, p. 407), and in some collectivist societies, "explanations of behavior seem to require an analysis of social roles, obligations, and situational factors" (Fiske et al., 1998, p. 915).

These well-documented cultural differences (Lillard, 1998; Lavelle, 2021) matter for XAI development. We do not challenge that AI programmers or other expert AI users need to have insights into a system's internals to debug it and so may prefer internalist XAI outputs (Bhatt et al., 2020). However, the findings just outlined cast doubts on the common view in XAI research that internalist explanations are analogous to how people *in general*, including lay users, preferentially explain behavior. The findings raise the possibility that potentially many individuals from collectivist cultures (which form 70% of the world population; Triandis, 1995) may often prefer or even require externalist explanations, i.e., explanations with more reference to context, social functions, norms, or others' behavior than to internal states. If XAI systems produce predominantly only internalist explanations and do not sufficiently cite external factors, people in collectivist cultures may find them unsatisfactory and less trustworthy.

To make the difference between internalist and externalist explanations with respect to XAI outputs more concrete, Table 2 provides four examples of potential externalist counterparts to the internalist XAI outputs from Table 1.

| |
|---|
| Internalist XAI: "I am $C x(\ )$ confident that $y$ will be correct based on $|S|$ past cases deemed similar to $x$." <br> Externalist XAI: "$Y$ will be correct because my task is to find the most likely result based on $|S|$ past cases deemed similar to $x$." |
| Internalist XAI: "Here is why the classifier thinks so [presentation of (e.g.) a decision tree]." <br> Externalist XAI: "The classifier produced this output because classification rules specify that given $x$, $y$ is the case." |
| Internalist XAI: "Why this exercise? Wiski thinks your current level matches that of this exercise!" <br> Externalist XAI: "Why this exercise? In most Wiski users with your current level, this level matched that exercise!" |
| Internalist XAI: "ShapeBot knows this is a [AI output] because ShapeBot realizes [decision factors]." <br> Externalist XAI: "ShapeBot classifies this as a [AI output] because ShapeBot's task is to do so when presented with [decision factors]." |

Table 2: Pairs of internalist and potential externalist XAI outputs

We currently lack the data to tell whether people from collectivist or individualist cultures will indeed respond differently to such XAI outputs because to the best of our knowledge (and based on the corpus analysis we report below), this has not yet been investigated. Our point here is that given the evidence that we have from previous psychological studies, there is reason to believe that differential reactions to these two kinds of XAI outputs are likely to occur in many individuals of the relevant cultures. It would therefore be valuable if XAI researchers experimentally tested and compared users' responses to the outlined internalist and externalist outputs. Given the current absence of explicit testing for or reflection on these potential differences, the popular use of "algorithm-centered" internalist post-hoc explanations in XAI developments (e.g., De Graaf & Malle, 2017; Zerilli et al., 2019; Zerilli, 2022; Ehsan et al., 2021) suggests that many XAI designs implicitly and problematically assume that Western explanatory needs and preferences are shared cross-culturally, revealing a cultural bias.

There are other cultural differences in human explanations and related cognitive processes than the individualist/internalist and collectivist/externalist variation. To draw XAI researchers' attention to them, in Table A1 in the Appendix, we present a range of psychological studies and reviews that we have not yet mentioned here and that strike us as especially relevant for XAI research. For instance, in experimental settings, participants from East Asia preferred more detailed explanations (Klein et al., 2014), indirect, contextualized communication (Wang et al., 2010), and similarity-based object categorization than Western participants did (Nisbett et al, 2001).

All of that said, classifying cultures as individualist and collectivist or as WEIRD and non-WEIRD may not be the best way to account for cultural variations in explanation because this approach risks homogenizing and stereotyping users from the related countries. To mitigate this, XAI researchers may refrain from these dichotomies and instead investigate more broadly where users from different cultural backgrounds are satisfied with one type of explanation, in which cases they may require or prefer internalist versus externalist outputs, or whether their choice is application dependent. We do not intend the individualist/collectivist and WEIRD/non-WEIRD categories to be definitive of a culture (e.g., WEIRD and non-WEIRD groups are heterogeneous, not always clearly distinct, and should not be reified; Ghai, 2021). We only employ these categories here because they offer interpretative tools for examining cross-cultural differences that have already been used insightfully in other AI-related research (e.g., differences in algorithmic aversion; Liu et al., 2023) and do capture reliable (but fluid) cultural differences in human explanations between some members of WEIRD and non-WEIRD populations, making them relevant for XAI and HHAI research.

To what extent are XAI researchers aware of the outlined cultural variations? To find out, we systematically reviewed XAI user studies.

## 3. A Systematic Analysis of XAI User Studies

Adapting key components from the Preferred Reporting Items for Systematic Reviews and Meta-Analyses (PRISMA) framework (Moher et al., 2009) and following a protocol used in previous work (Peters & Carman, 2023; Peters & Lemeire, 2023), we reviewed XAI user studies to answer four research questions (RQ):

*RQ1*. Do researchers that conduct XAI user studies indicate awareness that cultural variations may affect the generalizability of their results?
*RQ2*. What is the cultural background of the samples that XAI user researchers test?
*RQ3*. Do XAI researchers restrict their user study conclusions to their participants or study population, or generalize beyond them?
*RQ4*. Is the scope of researchers' conclusions related to the cultural diversity of their samples such that studies with broader conclusions are associated with more diverse samples?

## 3.1 Methodology

To identify relevant papers, in July 2022, we searched three major databases covering computer science and AI literature, i.e., Scopus, Web of Science, and arXiv, using a query containing 15 variants of key words related to XAI and end-users (for details, see Table A2, Appendix). The results were 2523 papers. After removing duplicates ($n = 535$), 1988 papers remained. Their titles and abstracts were scanned to find papers that met our selection criteria.

*Selection criteria.* We included any primary study (article, conference paper, chapter) that surveyed people on AI-based explanations of AI decisions and was published between January 2012 and July 2022. We excluded reviews, surveys, theoretical (incl. philosophical) papers, unpublished drafts (vs., e.g., arXiv preprints), guidelines, position papers, tutorials, technical, or applied papers (e.g., only introducing new XAI models), studies or surveys on other AI features than specific XAI outputs (e.g., 'algorithmic aversion'), and small-scale stakeholder or user studies with $\leq 5$ participants, which is too small a sample to ensure robust generalizations (Cazañas et al., 2017). We also excluded non-English papers. Of the 1988 papers, 192 remained for further screening, during which forward snowballing produced 14 more papers, resulting in 206 articles for full-text analysis (Figure 2).

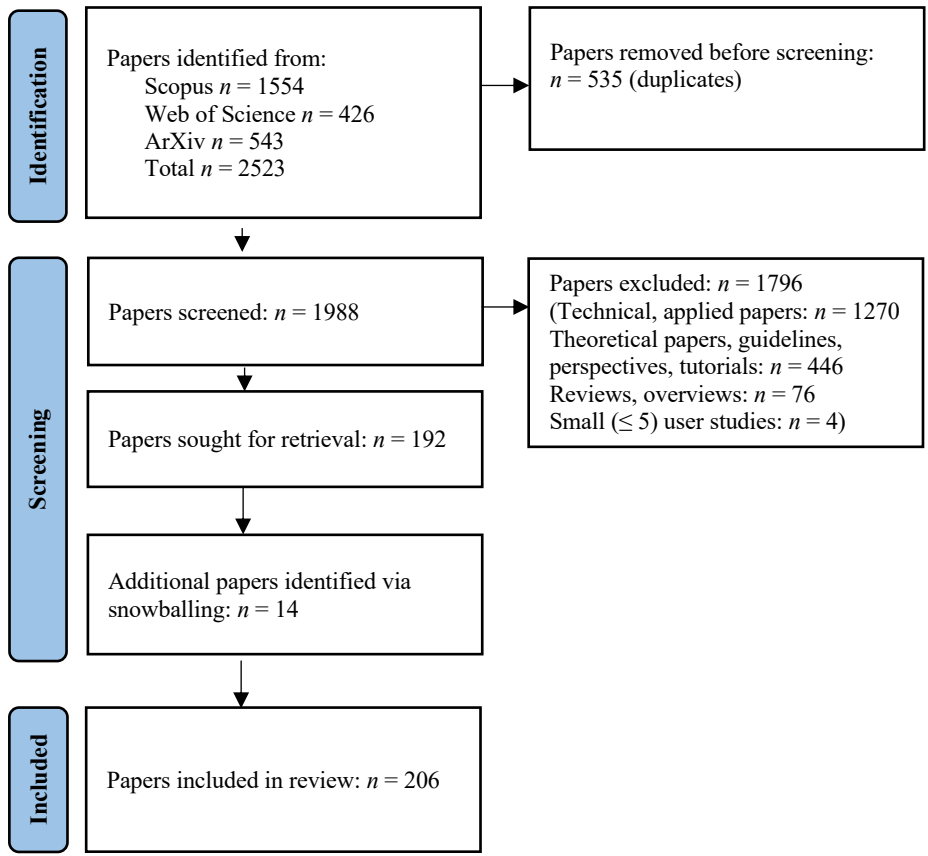

Figure 2: PRISMA flowchart of the systematic review

*Data extraction.* During full-text analysis, we (two researchers) independently classified papers by using pre-specified criteria (and a binary label, 0 = no; 1 = yes) to extract the following information. Apart from publication year and participant recruitment practice (e.g., conventional sampling or Amazon Mechanical Turk (MTurk) crowdsourcing), we extracted XAI output type, classifying papers as 'internalist' when they tested XAI explanations purporting to capture models' internal decision parameters (e.g., local feature importance), or as 'externalist' when they tested XAI explanations citing external factors (e.g., context, cultural norms, social situation) or involved XAI-user interaction (e.g., follow-up questions). Additionally, we classified papers on whether authors indicated awareness that culture can influence people's responding to XAI outputs in ways affecting the study generalizability.

We also extracted participants' cultural background, operationalizing it as participants' country or region (e.g., Europe) (Sawaya et al., 2017). Nationality or region is not always coextensive with cultural background (Taras et al., 2016). But it was typically the only clue of cultural belonging in the papers, and analyses found that alternative social aggregates (e.g., ethnicity) contributed only negligible explained variance to that already captured by nations (Akaliyski et al., 2021). Depending on the sample's country or region, we also labelled a paper as 'WEIRD', 'non-WEIRD', or 'mixed' based on previous studies' geographical categorizations (Klein et al., 2018; Yilmaz & Alper, 2019).

Finally, we identified an article's scope of conclusion based on the population to which results were generalized. Scientists commonly distinguish three types of populations: the *target population*, i.e., people to whom results are intended to be applied in real-world contexts (e.g., all users of a system X); the *study population*, i.e., users who are available and eligible for the study (e.g., US users meeting specific inclusion criteria); and the *study sample*, i.e., participants drawn from the study population (Banerjee & Chaudhury, 2010). We coded articles as 'restricted' if, throughout their text, authors did not extrapolate their findings beyond their sample or study population but instead used qualifiers (e.g., 'our participants'), quantifiers (e.g., 'some European users'), or past tense to limit their claims or recommendations to these populations, or otherwise indicated that they are study, sample, context, or culture specific. Authors may in contrast also describe results by using *generic*s, i.e., unquantified generalizations that refer to whole categories of people not specific, explicitly quantified sub-sets of them (e.g., 'Users prefer $X$' vs. 'Many (US, 75%, Western, etc.) users prefer $X$'). Or they may use other expressions that suggest that the results apply, for instance, to all non-experts, users, people, contexts, time, or cultures (for examples, see Table 3). If a paper had at least one such broad results claim in the mentioned sections, it was classified as 'unrestricted'. Papers with both restricted and unrestricted claims were also labelled 'unrestricted' because manuscripts are typically revised multiple times. If authors do not qualify their broader claims in the revisions, there is reason to believe they consider their broader generalizations warranted.

*Reliability.* For each classification, inter-rater agreement was calculated (Cohen's κ). It was consistently substantial (between κ = .71 and .90). We additionally asked two project-naïve researchers to independently classify the scope of conclusion variable for 25% of the data using our criteria. Inter-rater agreement between their and our classifications was also substantial (κ = .66 and .74, respectively), providing an additional reliability control for this variable. All remaining disagreements were resolved by discussion before the data were analyzed. All our data are publicly accessible on an OSF platform underline{here.}

## 3.2 Results

Most of the 206 XAI studies (94.7%, *n* = 195) in our sample were published between 2019 and 2022. Several studies used multiple recruitment practices, where 45.2% (*n* = 93) of all papers reported conventional sampling, followed by crowdsourcing via websites. The two most common websites were MTurk (29.1%, *n* = 60) and Prolific (9.2%, *n* = 19).

While some papers tested multiple kinds of XAI outputs, 88.8% (*n* = 183) of the papers focused on internalist explanations, and only 14.6% (*n* = 30) mentioned external factors (incl. XAI-user interaction) as relevant for users' perception of XAI outputs. Moreover, just 3.4% (*n* = 7) of the papers considered explanations that invoked external factors that may appear in collectivist explanations such as social rules, contexts, or social functions. None of the 206 papers explored potential differences in people's responding to internalist versus externalist XAI outputs that we outlined above.

*RQ1. Do researchers that conduct XAI user studies indicate awareness that cultural variations may affect the generalizability of their results?* 93.7% (*n* = 193) of the papers did not display any awareness (e.g., in discussion, limitation, or conclusion sections) that there may be cultural differences in how people perceive XAI outputs that can undermine broad extrapolations of results. Relatedly, these papers did not provide support (e.g., arguments or evidence) for the assumption that human explanatory needs are invariant across cultures.

*RQ2. What is the cultural background of the samples that XAI user researchers test?* 48.1% (*n* = 99) of the papers did not report cultural information about their samples. Across the remaining 107 papers, 32 countries or regions were mentioned. The three most frequent ones were the US (*n* = 53), the UK (*n* = 13) and Germany (*n* = 12) (for details, see Table A3, Appendix). Moreover, from the 107 papers, 81.3% (*n* = 87) had only WEIRD samples, exceeding the numbers of papers with mixed samples (10.3%, *n* = 11) and with only non-WEIRD samples (8.4%, *n* = 9).

*RQ3. Do XAI researchers restrict their user study conclusions to their participants or study population, or generalize beyond them?* Since 99 papers did not provide cultural information, they could have involved diverse samples. Broad generalizations may in this case be unproblematic. Since we could not determine cultural background in these papers, we analyzed only the remaining ones with this information (*n* = 107). 70.1% (*n* = 75) of them contained unrestricted conclusions, i.e., claims that suggested that the study results applied to all (e.g.) non-experts, people, users, consumers, humans, contexts, or time. Table 3 below presents examples (a full list of all unrestricted claims that we found in the papers can be found on our OSF platform here).

| |
|---|
| (1) "Our user study shows that non-experts can analyze our explanations and identify a rich set of concepts within images that are relevant (or irrelevant) to the classification process." (Schneider & Vlachos, 2023, p. 4196) |
| (2) "Our pilot study revealed that users are more interested in solutions to errors than they are in just why the error happened." (Hald et al., 2021, p. 218) |
| (3) "Our findings demonstrate that humans often fail to trust an AI when they should, but also that humans follow an AI when they should not." (Schmidt et al., 2020, p. 272) |
| (4) "We also found that users understand explanations referring to categorical features more readily than those referring to continuous features." (Warren et al., 2022, p. 1) |
| (5) "Both experiments show that when people are given case-based explanations, from an implemented ANN-CBR twin system, they perceive miss-classifications to be more correct." (Ford et al., 2020, p. 1) |
| (6) "Results indicate that human users tend to favor explanations about policy rather than about single actions." (Waa et al., 2018, p. 1) |
| (7) "Our findings suggest that people do not fully trust algorithms for various reasons, even when they have a better idea of how the algorithm works." (Cheng et al., 2019, p. 10) |

Table 3: Examples of unrestricted conclusions

*RQ4. Is the scope of XAI researchers' conclusions related to the cultural diversity of their samples such that studies with broader conclusions are associated with more diverse samples?* To address this question, focusing only on the papers with cultural information ($n = 107$), we first analyzed the scope of conclusions in the papers with only WEIRD, only non-WEIRD, and mixed samples. If studies with broader conclusions have more diverse samples, then one would predict that papers with unrestricted conclusions tend to have mixed samples, i.e., not either only WEIRD, or only non-WEIRD samples. Table 4 presents the comparisons. Unlike predicted, 90.7% ($n = 68$) of the 75 papers with unrestricted claims had in fact only WEIRD (84%, $n = 63$) or only non-WEIRD (6.7%, $n = 5$) samples. Moreover, if papers with broader conclusions had more diverse samples, then papers with unrestricted claims should include a higher proportion of papers with mixed samples compared to papers with restricted claims. However, a $\chi^2$ test showed that there was no evidence of a statistically significant difference of this kind ($p = 0.51$).

| | Scope of conclusion | | |
| Sample background | Restricted papers | Unrestricted papers | Total |
|---|---|---|---|
| Only non-WEIRD | 4 | 5 | 9 |
| Mixed | 4 | 7 | 11 |
| Only WEIRD | 24 | 63 | 87 |
| *Total* | 32 | 75 | 107 |

Table 4: Distribution of papers with unrestricted and restricted conclusions by sample composition

Furthermore, when relating the number of countries or regions sampled in each paper (which ranged from 1 to 19) to the scope of conclusion variable, we found that 82 papers mentioned only one country or region but nonetheless constituted 74.7% ($n = 56$) of all papers with unrestricted conclusions ($n = 75$) (Table A4, Appendix). Finally, to statistically analyze whether papers with unrestricted conclusions had more diverse samples than papers with restricted conclusions, we also conducted a Mann-Whitney $U$ test (as our data were not normally distributed) with the number of countries/regions as our dependent scale variable and scope of conclusions as the categorical independent variable. We found no evidence that unrestricted papers had or were correlated with a statistically significant higher number of countries or regions in their samples than the restricted papers ($p = 0.59$).

Our findings indicate significant shortcomings in many XAI user studies. But before interpreting the results, it is worth exploring whether researchers who have conducted literature reviews of XAI user studies have noticed any of the issues that we have just reported, i.e., a lack of awareness of relevant cultural variations, pervasive WEIRD sampling, or broad generalizations of XAI user study results from WEIRD to non-WEIRD populations. We therefore extended our systematic review to recent literature reviews of XAI user studies themselves.

## 4. A Meta-review of Reviews about XAI User Studies

Following the same procedure as before, we explored four questions:

*RQ1.* Do literature reviews about XAI user studies indicate that there may be cultural variations in explanatory needs that can affect the generalizability of study results?
*RQ2.* Do these reviews comment on WEIRD population sampling in XAI user studies?
*RQ3.* Do they comment on potential hasty generalizations in these studies?

*RQ4.* Do the authors of reviews about XAI user studies restrict their own conclusions from these studies to particular samples or study populations or generalize beyond them?

## 4.1 Methodology

To find reviews to analyze, in September 2022, we used the same three databases (Scopus, Web of Science, arXiv) and search strings as before but now added the specific restrictor "review" (for details, see Table A4, Appendix). The search results were 130 papers. 10 duplicates were removed. Titles and abstracts of the remaining 120 papers were scanned for articles meeting our inclusion criteria. We included any literature review of XAI user studies that was published between January 2012 and September 2022. We excluded any theoretical paper about AI principles or XAI guidelines, and any review paper about AI or XAI that did not focus on XAI user studies. Non-English publications were also excluded. 24 articles remained for further screening, during which forward snowballing produced 10 more papers, yielding 34 articles for full-text analysis (see Figure 4).

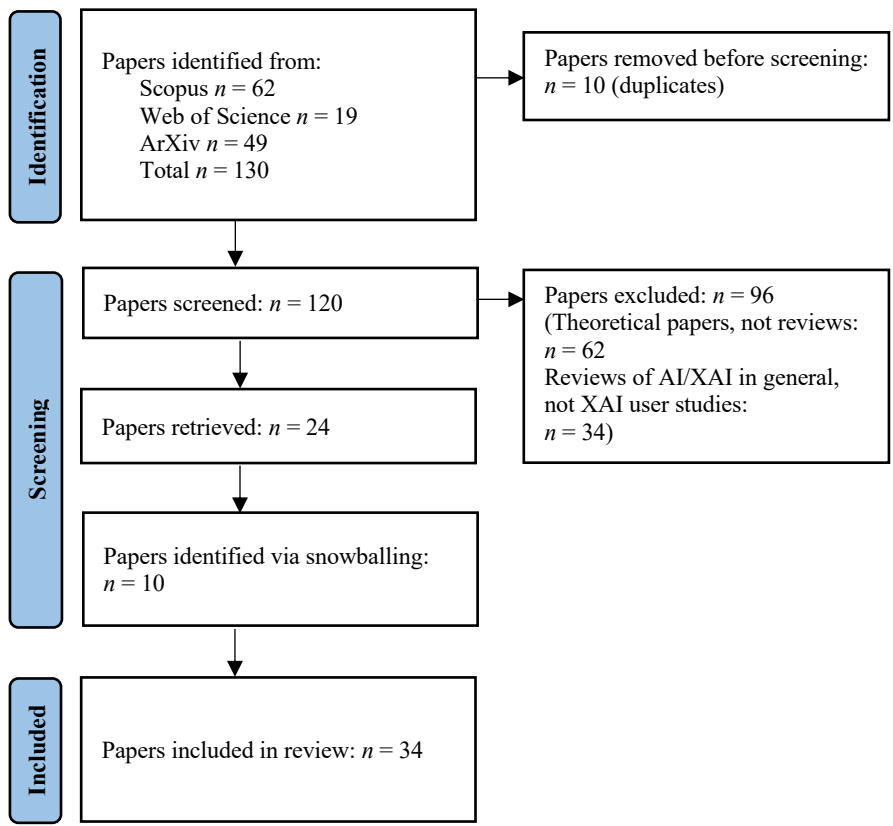

Figure 4: PRISMA flow diagram for the meta-review

During full-text analysis, we independently classified papers according to the following information (in addition to publication year). We applied a binary label (0 = no or 1 = yes) to a review if it (1) indicated that there might be cultural, contextual, or social variations in people's perceptions of XAI outputs that are relevant for user research, (2) contained comments on WEIRD sampling in XAI studies, (3) noted the relevance of stating cultural, national, or regional background in user studies, or (4) commented on potential hasty generalizations of results in these studies. We also classified reviews according to the scope of conclusions that they drew from XAI user studies, employing the same restricted versus unrestricted distinction as previously. To ensure reliability for the classifications, we again calculated inter-rater agreement. It was consistently substantial (between κ = .78 and κ = 94).

## 4.2 Results

In our sample of 34 literature reviews, most reviews (91.2%, *n* = 31) were published between 2019 and 2022 (Table A5, Appendix)

*RQ1. Do literature reviews about XAI user studies indicate that there may be cultural variations in explanatory needs that can affect the generalizability of study results?* In 82.4% (*n* = 28) of the reviews, this did not happen. Moreover, from the 6 reviews that briefly referred to culture, just 2 mentioned variations relevant for XAI, noting that there are "clear cultural differences in preference for simple versus complex explanations" (Mueller et al., 2019, p. 77) and "differences in the preference for personalized explanations depending on their cultural background" (Sperrle et al., 2020, p. 5). However, neither paper elaborated on or offered an overview of XAI-relevant cultural differences.

*RQ2. Do reviews about XAI user studies comment on WEIRD population sampling in these studies?* 94.1% (*n* = 32) of the reviews did not do so. Only 2 reviews (Sperrle et al., 2020; Laato et al., 2022) displayed sensitivity to the relevance of being explicit about XAI users' cultural, national, or regional background in XAI user studies. However, these papers did not specifically review XAI user studies to explore the extent of WEIRD sampling, nor did they offer quantitative data on it.

*RQ3. Do reviews of XAI user studies comment on potential hasty generalizations in these studies?* Just 1 of 34 reviews mentioned unwarranted extrapolations, writing that "perhaps the greatest challenge in the study of HAI [human-AI] teams […] is simply resisting the urge to overgeneralize experimental results" (Zerilli et al., 2022, p. 7). However, the authors did not provide quantitative evidence of the extent of hasty, overly broad generalizations in XAI papers and did not consider XAI-relevant cultural differences, or WEIRD sampling. They also themselves overgeneralized some study results when writing, for instance, that "in very simple automated tasks involving a single person, people tend to distrust automated aids whose errors they witness, unless an explanation is provided" (ibid). This brings us to *RQ4*.

*RQ4. Do the authors of reviews about XAI user studies restrict their own conclusions from these studies to particular samples or study populations generalize beyond them?* From all 34 reviews, 82.4% (*n* = 28) involved generalizations beyond any particular sample or (e.g., national) study population. Table 5 presents examples.

| |
|---|
| (1) "Users who are expert or self-confident in tasks that have been delegated to automation tend to ignore machine advice […]." (Zerilli et al., 2022, p. 4) |
| (2) "It has been recognised in the literature that counterfactuals tend to help humans make causal judgments." (Chou et al., 2022, p. 42) |
| (3) "Users tend to anthropomorphize AI and may benefit from humanlike explanations." (Laato et al., 2022, p. 14) |
| (4) "The textual explanations generated with GRACE were revealed to be more understandable by humans in synthetic and real experiments." (Islam et al., 2022, p. 20) |
| (5) "Humans lose trust in the explanation when soundness is low." (Gerlings et al., 2021, p. 5) |
| (6) "People tend to prefer complex over simple explanations if they can see and compare both forms." (Mueller et al., 2019, p. 77) |
| (7) "People rarely expect an explanation that consists of an actual and complete cause of an event." (Miller, 2019, p. 3) |

Table 5: Seven examples of unrestricted conclusions in XAI reviews

## 5. General Discussion and Recommendations

Our analyses reveal significant methodological limitations in much of the currently available XAI user research. We briefly revisit the three main findings of our two reviews and introduce mitigation strategies for the problems that our results highlight.

*(1) Lack of sensitivity to cultural variations in explanatory needs*. In the first analysis, we found that almost 90% of the XAI studies that we reviewed focused only on internalist explanations. As argued in Section 2, these explanations may better align with WEIRD individuals' explanatory needs than with those of non-WEIRD people in collectivist cultures, who may prefer externalist explanations (Henrich et al., 2010; Lavelle, 2021). Externalist explanations involving factors often highlighted in collectivist cultures were only explored in less than 4% of all studies. Moreover, in about 90% of the papers, authors did not display any awareness of cultural differences such as those discussed in section 2 and outlined in Table A1. Our meta-review of literature reviews about XAI user studies additionally revealed that the vast majority of these reviews (> 80%) were not sensitive to cultural variations in people's explanatory needs either. These findings suggest that XAI researchers routinely overlooked potentially relevant cultural differences that can affect human-AI interactions.

To tackle these problems, we recommend that AI journals increase the cultural diversity of their reviewer pool to ensure viewpoint variation during manuscript evaluation (Linxen et al., 2021). Conference organisers, in turn, can use platforms such as OpenReview, which makes reviewer reports public thereby allowing for an extra level of accountability (Wang et al., 2021). We also recommend that the increasing emphasis in reviews of XAI work on relating psychological findings to XAI developments (e.g., Miller, 2019; Rong et al., 2022) be extended to include data on the cultural differences summarized in Section 2 and Table A1.

*(2) WEIRD sampling.* We found that non-WEIRD populations were rarely sampled in XAI user research. This finding matches results from studies that explored sampling in HCI and HRI and report that 73-75% of papers tested only WEIRD populations (Linxen et al., 2021; Seaborn et al., 2023). However, our results suggest that the problem may be worse in XAI research, as more than 80% of XAI papers with relevant information involved only WEIRD participants. The findings of our meta-review add further weight to this problem because almost all (94%) of the reviews in our sample overlooked the predominately WEIRD sampling in XAI user studies.

There may be explanations for why WEIRD populations are over-represented. WEIRD countries may be a key market for XAI and hybrid human-AI designs. However, even if that is so, diverse sampling can still be advantageous because a significant number of people in WEIRD countries have diverse, non-WEIRD background (e.g., people from China, India, South America living in the USA) (Budiman & Ruiz, 2021). XAI products tested on more diverse users may thus ultimately be more profitable even in WEIRD countries, as they can appeal to a wider market.

Another reason for the pervasive WEIRD sampling may be that XAI user studies are predominantly conducted in WEIRD countries and geographically diverse sampling can be complicated. However, experiments through online platforms (e.g., MTurk or Prolific) are often feasible and have wider reach, enabling more diverse sampling. 29.1% ($n = 60$) of the XAI studies we reviewed already used MTurk. Even then, though, caution is warranted, as research suggests that most MTurkers (80%) come from the USA (Keith & Harms, 2016). It is worth noting that comparative studies found that Prolific has more diverse, less dishonest, more attentive, and reliable participants, providing higher quality data than MTurk (Peer et al., 2017). Yet, we found that only 9.2% of the reviewed XAI papers used Prolific, suggesting that many XAI researchers may be unaware of these differences. Thus, while conventional sampling should not be abandoned (e.g., it may be needed to study people without computer access), we recommend that XAI researchers recruit via Prolific, or LabintheWild, a crowdsourcing platform specifically developed to tackle the WEIRD sampling problem (Reinecke & Gajos, 2015), rather than MTurk to increase cultural diversity in user studies.

We acknowledge, however, that testing culturally diverse groups can also come with conceptual challenges because culture itself can and has been defined in multiple ways (Baldwin et al., 2006), where different ways of operationalizing culture (e.g., ethnicity, values, collective traits, country of residence, citizenship, heritage, shared language; Taras et al., 2009, 2016) can make comparisons between XAI studies and assessments of appropriate levels of generalizations difficult. Many existing measures of culture draw on Hofstede's (1980) methodology and his self-report questionnaire containing items about individualism/collectivism, power distance, uncertainty avoidance, and masculinity/femininity (Taras et al., 2009). Since Hofstede's original questionnaire is perhaps too long for inclusion into XAI user studies (it contained 126 questions), for XAI studies investigating, for instance, individualist/collectivist differences, we recommend that researchers adapt the related items from this questionnaire, as it is validated. That said, Hofstede's theory and methodology have also been criticized for being overgeneralizing (McSweeney, 2002), leading some technology researchers to use nationality as a proxy for culture instead (Ur & Wang, 2013; Sawaya et al., 2017). To capture that culture is a multidimensional construct, XAI researchers may therefore refrain from any single definition of culture and instead individually measure (via self-report items) users' nationality, racial/ethnic background, country of residence, home language and the relevant aspect of Hofstede's construct and then conduct regression analyses to identify and report the strongest predictor of responses to XAI outputs. This can enable insights into culture-related variations and may allow for comparisons and extrapolations across social groups and XAI user studies without invoking simplistic characterizations of culture.

Finally, it is important to note that as many as 48.1% ($n = 99$) of the reviewed studies did not report cultural (country/region) information about their samples. While these studies may have involved diverse population samples, the absence of reporting suggests that this information was not considered relevant for replication or generalizability. Not reporting on cultural information may be justified within a given study. However, it could also reflect implicit assumptions and biases about whether findings from particular populations are more generalizable than findings from other populations (Cheon et al., 2020). We therefore recommend that researchers either report information about participants' cultural backgrounds in ways discussed above or provide "constraints on generality" statements, specifying the study population and the basis for believing that the sample is representative of it or broader populations (for guidance on these statements, see Simons et al., 2017; Linxen et al., 2021).

*(3) Hasty generalizations of XAI study results.* Most of the XAI studies we analyzed contained conclusions that presented findings as if they held for whole categories of people (e.g., experts, users, humans) even when they had only tested WEIRD populations or a single country. Generalizations from WEIRD to non-WEIRD populations need not be unwarranted. Researchers who produced such extrapolations might have had good grounds to assume that this particular dimension of demographic variation was irrelevant for their study. So, it does not follow from the evidence that XAI user researchers drew conclusions about populations much wider than their study population that these generalizations were unwarranted. However, if all the unrestricted conclusions we found had been based on researchers' reflection on relevant or irrelevant demographic differences, there should have been an indication of this reflection in their papers. This is because to fully establish a study conclusion and make the study reproducible, all underlying assumptions that justify the conclusions (including the potential assumption that people's explanatory needs or preferences are cross-culturally invariant) need to be made explicit. Yet, as noted, we found that more than 90% of the reviewed XAI user studies did not contain any evidence of reflection on XAI-relevant cultural differences or invariance. Furthermore, we could not find any evidence that XAI user studies with broader claims had or were associated with more diverse samples. Hence, the unrestricted conclusions in most of the reviewed XAI papers were hasty generalizations, i.e., claims whose scope was broader than warranted by the evidence and justification provided by the researchers. Since psychological research suggests that explanatory needs likely differ between WEIRD and non-WEIRD populations, as discussed in Section 2, the pervasive insufficiently supported extrapolations that we found from WEIRD samples to other populations may indicate a cultural "generalization bias" (Peters et al., 2022) toward WEIRD populations in many currently available XAI user studies.

Our findings fill a significant gap in previous studies in HCI and HRI that reported generalizability problems related to WEIRD sampling in these fields (Linxen et al., 2021; Seaborn et al., 2023). This is because these studies did not measure the scope of the generalizations that researchers produced thus leaving it unclear whether methodological shortcomings were involved. Indeed, while hasty generalizations across cultures have been reported in other fields (e.g., Peters & Lemeire, 2023) until now it has remained unknown whether they also occur in XAI, allowing XAI researchers to ignore or deny their presence in the field of XAI. Our results block this potential move, which is important, as encountering overgeneralizations in XAI user studies is particularly disconcerting. XAI user study results can directly feed into the production of XAI that a wide range of people later interact with (Ehsan et al., 2021; Ding et al., 2022; Okolo et al., 2022). These results can affect the way human-AI hybrid systems are developed via influencing what XAI models are included in HHAI designs. Generalizing results to cultural groups to whom they do not apply can hide that certain XAI and human-AI hybrid systems may only meet the explanatory needs of individuals with a particular cultural background, raising ethical concerns about both explainability and inclusivity.

We thus recommend that XAI studies be conducted in collaborations with researchers or participants from different cultures. To further mitigate hasty generalizations, XAI researchers should consider restricting their user study conclusions by using quantifiers ('US users', 'our participants', 'many users', etc.), qualifiers (e.g., 'may', 'can'), or past tense (Peters et al., 2022). Table A7 in the Appendix presents examples of restricted versions of the unrestricted conclusions from Table 3.

## 6. Limitations

There are several constraints on the generalizability of our own analysis results. First, our literature search was limited to three major databases of scientific literature covering XAI studies. Second, there may also be XAI user studies that do not use our specific search terms and that we may have overlooked. Third, we focused only on English publications on XAI user research and may have overlooked, for instance, recent research from non-Western institutions that was not published in English. Indeed, the current boom of AI research, particularly in China, may significantly counteract the cultural bias in XAI user studies that we reported (Min et al., 2023). Future research analyzing the cultural diversity and models employed in Chinese language XAI user studies are therefore desirable. However, since most scientific studies are now published in English (Ramírez-Castañeda, 2020), our findings remain important because of the size and influence of English within the scientific community. Another limitation of our analyses is that we used country or region as proxies for culture, as it was typically the only culture-related information in the reviewed papers. Using this proxy ignores expatriates, mixed national demographics, and shared, technology-facilitated experience. We therefore welcome future XAI research reviews that explore other proxies for cultural background in XAI user studies.

## 7. Conclusion

XAI systems play an increasingly more significant role in many human-AI interactions because they can make opaque AI models more trustworthy to people, facilitating human control over these models. XAI developers are thus doing important work that is directly relevant for hybrid human-AI (HHAI) systems. Here, we examined whether currently popular XAI systems for lay-users are equally suitable for people from different cultural backgrounds. We argued that XAI systems that produce internalist explanations (referring to mental states, e.g., beliefs) are currently popular but may cater primarily to the explanatory needs of people from individualist, typically WEIRD cultures. Psychological studies found that while most people from individualist cultures preferred human internalist explanations, people from collectivist, commonly non-WEIRD cultures tended to favor externalist explanations (referring to social roles, context, etc.). To help raise XAI and HHAI developers' awareness of these and other cultural variations relevant for XAI design and human-AI interactions, we provided a table offering an empirically informed overview of them (see Table A1 in the Appendix).

To support our claim that these variations are currently overlooked in XAI research, we analyzed 206 XAI user studies. Most of them contained no evidence that the researchers were aware of cultural variations in explanatory needs. Most studies also tested only WEIRD populations but researchers routinely generalized results beyond them. When we additionally analyzed 34 reviews of XAI user studies, we found that these problems went largely unnoticed even by most reviewers of these studies.

In offering evidence of XAI-relevant cultural variations, of a widespread oversight of them in the field of XAI, and of pervasive WEIRD sampling paired with extrapolations to non-WEIRD populations, this paper uncovers both a cultural bias toward WEIRD populations and an important knowledge gap in the field of XAI regarding how culturally diverse users may respond to widely used XAI systems. If human-AI hybrids include XAI systems of the kind tested in most of the user studies we reviewed then these hybrids may inherit the mentioned cultural bias and be less inclusive than they appear and could be. We hope that our analyses help stimulate cross- and multi-cultural XAI user studies and improve the vital work that XAI and HHAI developers are doing in making AI systems more explainable and useful for all stakeholders no matter their cultural background.

**Author statement**

UP conceived and designed the study, collected the data, did the data analysis, developed the argumentation, wrote the first draft, and did the editing.

MC assisted with the collection of the data, data classification, revising, and editing of the paper.

**Acknowledgements**

We would like to thank Apolline Taillandier and Charlotte Gauvry for cross-checking the classifications of some of our key data. We are also very grateful for helpful comments from Caroline Gevaert, Benjamin Rosman, Alex Krauss, and three reviewers of this journal. We do not have funding to declare, and we grant JAIR the permission to publish this paper.

# Appendix

| Author(s) | Country/region | Cultural variation | Relation to XAI |
|---|---|---|---|
| Howell (1981), Wierzbicka (1992), Lebra (1993) | Incl. Peru, Japan, US | Some non-Western cultures lack concepts comparable to Western psychological concepts 'think', 'belief', or 'desire' | Use of mentalistic XAI framing (e.g., 'AI thinks _') |
| Miller (1984), Al-Zahrani and Kaplowitz (1993), Morris and Peng (1994), Lee et al. (1996), Choi and Nisbett (1998) | Incl. India, US, Korea, China, Japan, Saudi Arabia | Non-Western (vs. Western) study participants referred to situations/social rules to explain behavior and were less susceptible to mistakenly explaining it through agents' internal states when situational causes were available | Preferences for internalist vs. externalist outputs |
| Choi et al. (2003) | Korea, US | When explaining behavior, Korean participants preferred more contextual information than their US counterparts | Preferences for XAI output scope |
| Klein et al. (2014) | Malaysia, US | Malaysian participants preferred more detailed explanations for indeterminate situations; US participants favored simpler ones | Preferences for XAI output complexity |
| Hall and Hall (1990), Sanchez-Burks et al. (2003), Wurtz (2005), Rau et al. (2009), Wang et al. (2010), Lee and Sabanović (2014) | High-context (e.g., Japan, China; South Korea), low-context (e.g., US, Germany) cultures | Participants from Western, low-context cultures (communication with low use of non-verbal cues) preferred direct, explicit communication (e.g., "Drink water."); participants from East Asian, high-context cultures preferred indirect, implicit communication (e.g., "Drinking water may alleviate headaches."), e.g., in robot recommendations | Preferences for XAI communication style |
| Nisbett et al. (2001), Norenzayan et al. (2002), Varnum et al. (2010), Henrich et al. (2010), Klein et al. (2018) | Western, East-Asian countries | Western participants displayed more analytic thinking (i.e., rule-based object categorization, context-independent understanding of objects, formal logic in reasoning); East-Asian participants displayed more holistic thinking (i.e., similarity-based object categorization, focus on context, intuition in reasoning) | Preferences for XAI output content (e.g., rule-based vs. example-based) |
| Otterbring et al. (2022) | East-Asian, US | East-Asian participants preferred abstract figures representing conformity; US participants favored objects representing uniqueness | Preferences for XAI output content and format |
| Reinecke and Gajo (2014), Alexander et al. (2021) | Incl. Russia, Macedonia, Australia, China, Saudi Arabia | Regarding websites' visual complexity/design attributes (layout, navigation, etc.), Australian users focused on textual items; Chinese users scanned the whole page | Preferences for XAI output complexity and format |
| Baughan et al. (2021) | US, Japan | Visual attention differences affected website search: Japanese participants remembered more and faster found contextual website information than US counterparts | Preferences for XAI output complexity and format |
| Van Brummelen et al. (2022) | US, Singapore, Canada, NZ, Indonesia, Iran, Japan, India | Non-WEIRD participants' perspectives emphasized virtual agent artificiality; WEIRD perspectives emphasized human-likeness | Social embedding can influence perceptions of AI |

Table A1: Select overview of psychological and HCI research with relevance for XAI design

**Scopus** (searched July 2022):

TITLE-ABS-KEY ("XAI" OR "Explainable AI" OR "transparent AI" OR "interpretable AI" OR "accountable AI" OR "AI explainability" OR "AI transparency" OR "AI accountability" OR "AI interpretability" OR "model explainability" OR "explainable artificial intelligence" OR "explainable ML" OR "explainable machine learning" OR "algorithmic explicability" OR "algorithmic explainability") AND ("end user" OR "end-user" OR "audience" OR "consumer" OR "user" OR "user study" OR "user survey" OR "developer") AND (LIMIT-TO (DOCTYPE,"cp") OR LIMIT-TO (DOCTYPE,"ar") OR LIMIT-TO (DOCTYPE,"ch")) AND (LIMIT-TO (PUBYEAR,2022) OR LIMIT-TO (PUBYEAR,2021) OR LIMIT-TO (PUBYEAR,2020) OR LIMIT-TO (PUBYEAR,2019) OR LIMIT-TO (PUBYEAR,2018) OR LIMIT-TO (PUBYEAR,2017) OR LIMIT-TO (PUBYEAR,2016) OR LIMIT-TO (PUBYEAR,2012)) AND (LIMIT-TO (LANGUAGE,"English"))

**Web of Science** (searched July 2022):

(ALL=("XAI" OR "Explainable AI" OR "transparent AI" OR "interpretable AI" OR "accountable AI" OR "AI explainability" OR "AI transparency" OR "AI accountability" OR "AI interpretability" OR "model explainability" OR "explainable artificial intelligence" OR "explainable ML" OR "explainable machine learning" OR "algorithmic explicability" OR "algorithmic explainability")) AND ALL=("end user" OR "end-user" OR "audience" OR "consumer" OR "user" OR "user study" OR "user survey" OR "developer") and Article or Proceedings Papers or Early Access or Book Chapters(Document Types) and English (Languages)

Refined by all 'Publication Years' (2012-01-01 to 2022-12-31)

**ArXiv** (searched July 2022):

Query: order: -announced_date_first; size: 50; date_range: from 2012-01-01 to 2022-12-31; classification: Computer Science (cs); include_cross_list: True; terms: AND all="XAI" OR "Explainable AI" OR "transparent AI" OR "interpretable AI" OR "accountable AI" OR "AI explainability" OR "AI transparency" OR "AI accountability" OR "AI interpretability" OR "model explainability" OR "explainable artificial intelligence" OR "explainable ML" OR "explainable machine learning" OR "algorithmic explicability" OR "algorithmic explainability"; AND all="end user" OR "end-user" OR "audience" OR "consumer" OR "user" OR "user study" OR "user survey" OR "developer"

Table A2: Systematic literature review search strings

| Country | N | Country | N | Country | N |
|---|---|---|---|---|---|
| No details | 99 | South/Latin America | 5 | Norway | 1 |
| US | 53 | Sweden | 4 | Denmark | 1 |
| UK | 13 | India | 4 | Iceland | 1 |
| Germany | 12 | Netherlands | 4 | South Korea | 1 |
| Canada | 9 | Asia | 3 | New Zealand | 1 |
| Europe | 9 | Switzerland | 3 | Portugal | 1 |
| Ireland | 7 | Belgium | 2 | Africa (unspecified) | 1 |
| North America | 6 | Finland | 2 | Americas (unspecified) | 1 |
| China | 5 | France | 2 | Rest of the world | 1 |
| Italy | 5 | Brazil | 2 | Russia | 1 |
| Australia | 5 | Japan | 2 | Costa Rica | 1 |

Table A3: Frequency of nationalities/regions in the reviewed XAI studies

| Countries or regions | Scope of conclusion: Restricted | Scope of conclusion: Unrestricted | Total |
|---|---|---|---|
| 1 | 26 | 56 | 82 |
| 2 | 2 | 5 | 7 |
| 3 | 2 | 5 | 7 |
| 4 | 0 | 3 | 3 |
| 5 | 0 | 1 | 1 |
| 6 | 2 | 3 | 5 |
| 8 | 0 | 1 | 1 |
| 19 | 0 | 1 | 1 |
| **Total** | 32 | 75 | 107 |

Table A4: Numbers of countries/regions in papers with restricted and unrestricted conclusions

| Year | Number |
|---|---|
| 2012-2016 | 0 |
| 2017 | 1 |
| 2018 | 2 |
| 2019 | 3 |
| 2020 | 6 |
| 2021 | 12 |
| Sept 2022 | 10 |
| **Total** | 34 |

Table A5: Reviews of XAI user study papers per year

**Scopus** (searched September 2022):

TITLE-ABS-KEY ( "XAI" OR "Explainable    AI" OR "transparent    AI" OR "interpretable AI" OR "accountable    AI" OR "AI    explainability" OR "AI    transparency" OR "AI accountability" OR "AI interpretability" OR "model explainability" OR "explainable artificial intelligence" OR "explainable    ML" OR "explainable    machine    learning" OR "algorithmic explicability" OR "algorithmic        explainability" ) AND ( "end        user" OR "end-user" OR "audience" OR "consumer" OR "user" OR "user            study" OR "user survey" OR "developer" ) AND ( LIMIT-TO ( PUBYEAR , 2022 ) OR LIMIT-TO ( PUBYEAR , 2021 ) OR LIMIT-TO ( PUBYEAR , 2020 ) OR LIMIT-TO ( PUBYEAR , 2019 ) OR LIMIT-TO ( PUBYEAR , 2018 ) OR LIMIT-TO ( PUBYEAR , 2017 ) OR LIMIT-TO ( PUBYEAR , 2016 ) OR LIMIT-TO ( PUBYEAR , 2012 ) ) AND ( LIMIT-TO ( DOCTYPE , "re" ) ) AND ( LIMIT-TO ( LANGUAGE , "English" ) )

**Web of Science** (searched September 2022):

(ALL=("XAI" OR "Explainable AI" OR "transparent AI" OR "interpretable AI" OR "accountable AI" OR "AI explainability" OR "AI transparency" OR "AI accountability" OR "AI interpretability" OR "model explainability" OR "explainable artificial intelligence" OR "explainable ML" OR "explainable machine learning" OR "algorithmic explicability" OR "algorithmic explainability")) AND ALL=("end user" OR "end-user" OR "audience" OR "consumer" OR "user" OR "user study" OR "user        survey"        OR        "developer")        and Review        Article (Document Types) and English (Languages)

Refined by all 'Publication Years' (2012-01-01 to 2022-12-31)

**ArXiv** (searched September 2022):

Query: order: -announced_date_first; size: 200; date_range: from 2012-01-01 to 2022-12-31; classification: Computer Science (cs); include_cross_list: True; terms: AND all="XAI" OR "Explainable AI" OR "transparent AI" OR "interpretable AI" OR "accountable AI" OR "AI explainability" OR "AI transparency" OR "AI accountability" OR "model explainability" OR "explainable artificial intelligence" OR "explainable ML" OR "explainable machine learning" OR "algorithmic explicability" OR "algorithmic explainability"; AND all="end user" OR "end-user" OR "audience" OR "consumer" OR "user" OR "user study" OR "user survey" OR "developer"; AND all="review"

Table A6: Systematic meta-review search strings

| |
|---|
| (1) "Our user study shows that non-experts can analyze our explanations and identify a rich set of concepts within images that are relevant (or irrelevant) to the classification process." (Schneider & Vlachos, 2023, p. 4196)<br>*Restricted:* "Our user study shows that non-expert **participants could** analyze our explanations and identify a rich set of concepts within images that **were** relevant (or irrelevant) to the classification process." |
| (2) "Our pilot study revealed that users are more interested in solutions to errors than they are in just why the error happened." (Hald et al., 2021, p. 218)<br>*Restricted:* "Our pilot study revealed that users **were** more interested in solutions to errors than they **were** in just why the error happened." |
| (3) "Our findings demonstrate that humans often fail to trust an AI when they should, but also that humans follow an AI when they should not." (Schmidt et al., 2020, p. 272)<br>*Restricted:* "Our findings demonstrate that **participants** often fail**ed** to trust an AI when they **should have trusted it**, but also that they follow**ed** an AI when they should not **have done** so." |
| (4) "We also found that users understand explanations referring to categorical features more readily than those referring to continuous features." (Warren et al., 2022, p. 1)<br>*Restricted:* "We also found that users **understood** explanations referring to categorical features more readily than those referring to continuous features." |
| (5) "Both experiments show that when people are given case-based explanations, from an implemented ANN-CBR twin system, they perceive miss-classifications to be more correct." (Ford et al., 2020, p. 1)<br>*Restricted:* "Both experiments show that when **participants were** given case-based explanations, from an implemented ANN-CBR twin system, they perceiv**ed** miss-classifications to be more correct." |
| (6) "Results indicate that human users tend to favor explanations about policy rather than about single actions." (Waa et al., 2018, p. 1)<br>*Restricted:* "Results indicate that **participants** tend**ed** to favor explanations about policy rather than about single actions." |
| (7) "Our findings suggest that people do not fully trust algorithms for various reasons, even when they have a better idea of how the algorithm works." (Cheng et al., 2019, p. 10)<br>*Restricted:* Our findings suggest that people **did** not fully trust algorithms for various reasons, even when they **had** a better idea of how the algorithm works." |

Table A7: Restricted versions of unrestricted conclusions. The parts in bold are the restricting components.